\documentclass{elsart}
\usepackage{amssymb}
\usepackage[dvips]{graphicx}
\usepackage{color}
\def\dag{\dagger}
\def\ndag{\mbox{}}
\def\ed{\varepsilon_{d}}
\def\ek{\varepsilon_{k}}

\begin{document}
\begin{frontmatter}
\title{Thermoelectric effects in quantum dots}
\author[igce]{M.Yoshida\corauthref{cor}};
\corauth[cor]{Instituto de Geoci\^encias e Ci\^encias Exatas de Rio Claro, UNESP, Brazil}
\ead{myoshida@rc.unesp.br} 
\author[ifsc]{L.N. Oliveira}
\address[igce]{Unesp-Universidade Estadual Paulista, Departamento de
  F\'{\i}sica, IGCE-Rio Claro, Brazil}
\address[ifsc]{Instituto de F\'{\i}sica de S\~ao Carlos, USP, Brazil}

\begin{abstract}
We report a numerical renormalization-group study of the
thermoelectric effect in the single-electron transistor (SET) and
side-coupled geometries. As expected, the computed thermal conductance and
thermopower curves show signatures of the Kondo effect and of Fano
interference. The thermopower curves are also affected by
particle-hole asymmetry.
\end{abstract}

\end{frontmatter}
\section{Introduction}
The transport properties of mesoscopic devices are markedly affected
by electronic correlations. Gate potentials applied to such devices
give experimental control over effects once accessible only in special
arrangements. In particular, the Kondo effect and Fano anti-resonances
have been unequivocally identified in the conductance of
single-electron transistors (SET) \cite{gold1,gold2,gores}; of
Aharonov-Bohm rings \cite{mahalu}; and of quantum wires with
side-coupled quantum dots \cite{sato}. Another achievement was a recent
study of the thermopower, a quantity sensitive to particle-hole
asymmetry that monitors the flux of spin entropy \cite{scheibner}. This
work presents a numerical renormalization-group
study \cite{wilson,mak,viv} of the thermoelectric properties of
nanodevices. We consider a quantum dot coupled to conduction electrons
in the two most widely studied geometries: the single-electron
transistor (SET), in which the quantum dot bridges two-dimensional gases
coupled to electrodes; and the $T$-shaped device, in which a quantum
dot is side-coupled to a quantum wire.

\section{Thermoelectric properties}
Thermoelectric properties are traditionally studied in two
arrangements: the Seebeck (open circuit) and Peltier (closed circuit)
setups \cite{ziman}. In the former, the steady-state electric current
vanishes. A temperature gradient drives electrons towards the coldest
region, and induces an electric potential difference between the hot
and the cold extremes. The expression $S= - \Delta V/\Delta T$, where
$\Delta V$ is the potential difference induced by the temperature
difference $\Delta T$, then determines the thermopower $S$. Since the
electrons transport heat, the heat current $Q$ can also be measured,
and the thermal conductance $\kappa$ can be obtained from the relation
$ Q = -\kappa \nabla T$.

In the Peltier setup, a current $J$ is driven through a circuit
kept at uniform temperature. The heat flux $Q =\Pi J$ is then measured and determines 
the Peltier coefficient $\Pi$, which is proportional to the thermopower: $\Pi = S T$. 

We prefer the Seebeck setup. The transport coefficients are then
computed from the integrals \cite{kim}
\begin{equation}\label{eq:1}
I_{n}(T) = -\frac{2}{h} \int_{-D}^{+D} \,\, \varepsilon^{n}\,\,
\frac{\partial f(\varepsilon)}{\partial \varepsilon}\,\, {\mathcal
  T}(\varepsilon,T)\,\, d\varepsilon\qquad(n=0,1,2).
\end{equation}
where ${\mathcal T}(\varepsilon,T)$ is the transmission probability
at energy $\epsilon$ and temperature $T$, $f(\varepsilon)$ is the Fermi
distribution and $D$ is the half width of the conduction band.  The
electric conductance $G$, the thermopower $S$, and the thermal
conductance $\kappa$ are given by \cite{kim}
\begin{eqnarray}
G &=& e^{2} I_{0}(T) \\ S &=& - \frac{I_{1}(T)}{e T I_{0}(T)}
\\ \kappa &=& \frac{1}{T} \left\{I_{2}(T) -
\frac{I_{1}^{2}(T)}{I_{0}(T)}\right\},
\end{eqnarray}
respectively. 
Our problem, therefore, is to compute ${\mathcal T}(\varepsilon,T)$
for a correlated quantum dot coupled to a gas of non-interacting electrons.

\section{Thermal Conductance and Thermopower of a SET}
Recent experiments \cite{gores} have detected Fano anti-resonances in
coexistence with the Kondo effect in SETs. The interference indicates
that the electrons can flow through the dot or tunnel 
directly from on electrode to the other. The transport properties of
the SET can be studied by a modified Anderson model \cite{hofst,bulka},
which in standard notation is described by the Hamiltonian
\begin{eqnarray}
H &=& \sum_{k,\alpha} \ek c_{k,\alpha}^{\dag}
c_{k,\alpha}^{\ndag} +
t\sum_{k,k'}(c_{k,L}^{\dag}c_{k,R}^{\ndag} +
H.c.)+ V \sum_{k,\alpha}
(c_{k,\alpha}^{\dag} c_{d}^{\ndag} + H.c. ) + H_{d}.\label{eq:ham}
\end{eqnarray}
Here the quantum-dot Hamiltonian is $H_{d} = \ed
c_{d}^{\dag}c_{d}^{\ndag} + U n_{d,\uparrow} n_{d,\downarrow}$, with a
dot energy $\ed$, controlled by a gate potential applied to the dot,
that competes with the Coulomb repulsion $U$.  The summation index $\alpha$ on
the right-hand side takes the values $L$ and $R$, for the left and right
electrodes respectively. The tunneling amplitude $t$ allows transitions between the
electrodes, while $V$ couples the electrodes to the quantum dot. The
Hamiltonian~(\ref{eq:ham}) being invariant under inversion, it is
convenient to substitute even ($+$) and odd ($-$) operators
$c_{k\pm}=(c_{kR}\pm c_{kL})/\sqrt2$ for the $c_{kL}$ and $c_{kR}$. It
results that only the $c_{k+}$ are coupled to the quantum dot. For
brevity, we define the shorthand $\gamma\equiv\pi\rho t$, where $\rho$
is the density of conduction states; and the dot-level width $\Gamma
\equiv \pi\rho V^2$.
\begin{figure}[t]
\begin{center}
\includegraphics[clip=true,width=8.5cm]{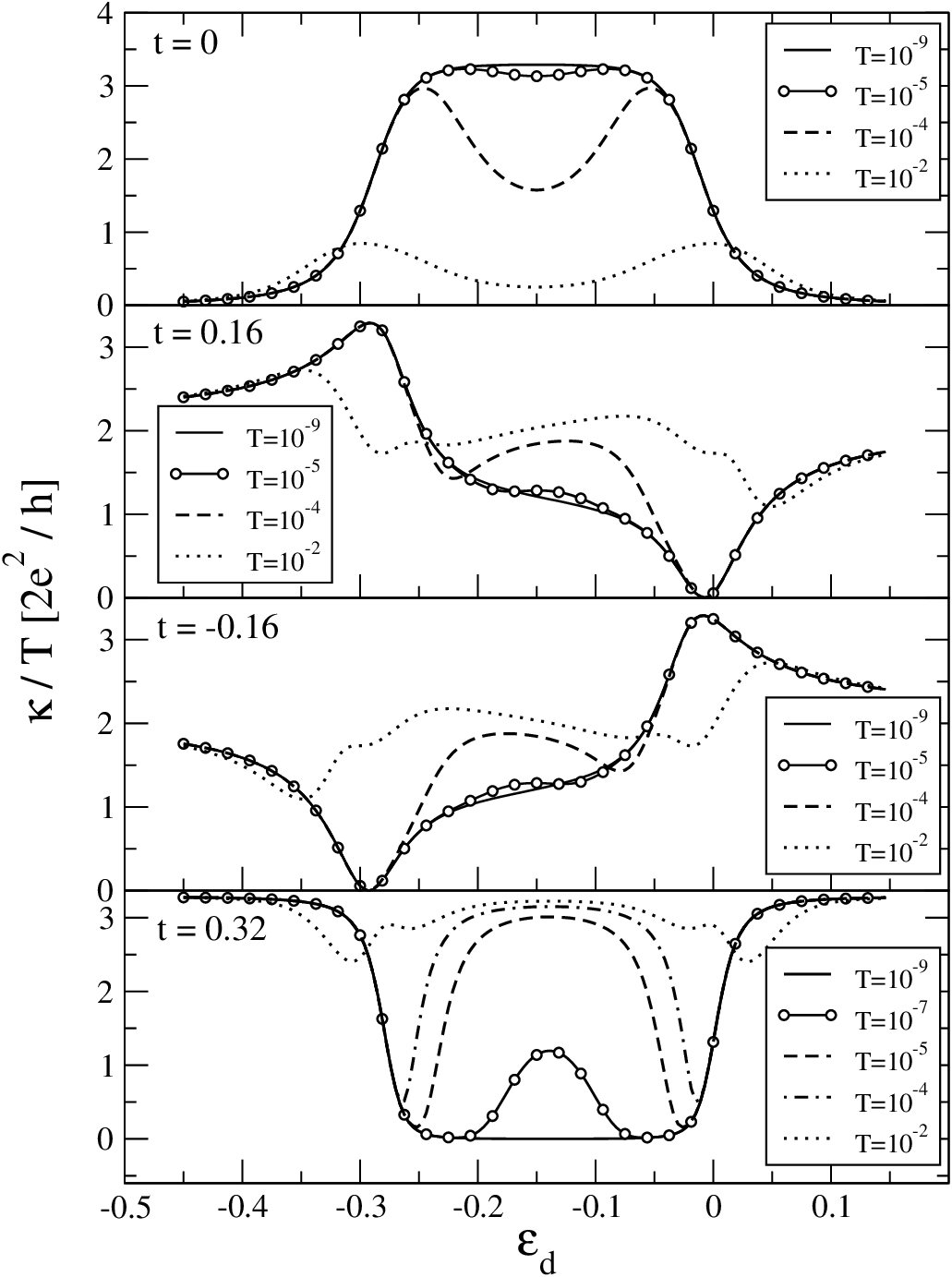}
\end{center}
\caption{{Thermal conductance $\kappa$, normalized by the
    temperature $T$, as a function of the dot energy for $U=0.3\,D$
    and four $t$s, at the indicated temperatures. The top panel, with
    $t=0$, shows no sign of interference. The second (third) panel,
    with $t=0.16\,D$ ($t=-0.16\,D$) displays a Fano antiresonance. In
    the bottom panel, $t=0.32\,D$, the conductance vanishes in the
    Kondo valley as $T\to0$.}\label{fig:1}}
\end{figure}

{In the absence of magnetic fields, the transmission
  probability through the SET is \cite{kim1,hofst}
\begin{eqnarray}\label{eq:transmit}
{\mathcal T}(\varepsilon,T)&=&T_{0} + \frac{4\Gamma
  \sqrt{T_{0}R_{0}}}{1+\gamma^{2}} {\Re}\{G_{d,d}\}(\varepsilon,T) + \nonumber
\frac{2\Gamma(T_{0}-R_{0})}{1+\gamma^{2}}{\Im}\{G_{d,d}\}(\varepsilon,T),
\end{eqnarray}
where $G_{dd}(\epsilon, T)$ is the retarded Green's function for the
dot orbital, and we have defined $T_0\equiv 4\gamma^{2}/(1+\gamma^{2})^{2}$
and $R_0\equiv1-T_0$.

To compute $\mathcal{T}$, we rely on the numerical-renormalization group (NRG)
diagonalization of the model Hamiltonian \cite{wilson}. Although the resulting
eigenvectors and eigenvalues yield essentially exact results for $\Im
\{G_{dd}\}(\epsilon,T)$, the direct computation of $\Re\{G_{dd}\}$ is
unwieldy. We have found it more convenient to define the Fermi
operator
\begin{eqnarray}
b \equiv \left(\frac{2 \Gamma}{1+\gamma^{2}}\right)^{1/2}
c_{d} + \left(\frac{4\gamma^ {2}}{1+\gamma^{2}}\right)^{1/2}\frac1{\sqrt{\pi\rho}}
\sum_k c_{k+},
\end{eqnarray}
because the imaginary part of its retarded Green's function
$G_{bb}(\epsilon,T)$ is directly related to the transmission
probability: aided by the two equations of motion relating $G_{kk'}$
to $G_{dk}$, and $G_{kd}$ to $G_{dd}$, straightforward manipulation of
Eq.~(\ref{eq:transmit}) show that ${\mathcal T}(\varepsilon,T) = - \Im
\{G_{b,b}\}(\varepsilon,T)$. In practice, we (i) diagonalize $H$
iteratively \cite{wilson}; (ii) for each pair of resulting eigenstates
$(|m\rangle, |n\rangle$), compute the matrix elements $\langle
m|b_{\sigma}|n\rangle$ ; (iii) thermal average the results
\cite{wanda,sandra} to obtain ${\Im}\{G_{b,b}\}(\varepsilon,T)$;
(iv) substitute the result for $\mathcal {T}(\varepsilon,T)$ in
Eq.~(\ref{eq:1}); and (v) evaluate the integral for $n=0,1,2$ to obtain $I_n(T)$
($n=0, 1, 2$).

Figure~\ref{fig:1} shows numerical results for the thermal conductance
as a function of the gate energy $\varepsilon_{d}$. Well above or
  well below the Kondo temperature $T_K$, we expect the thermal and
electric conductances to obey the Wiedemann-Franz law $\kappa/T =
\pi^{2} G/3$ and hence show the thermal conductance normalized by the
temperature $T$.  Each panel represents a tunneling parameter $t$ and
displays the thermal-conductance profile for the indicated
temperatures.  All curves were computed for $\Gamma= 10^{-2} D$, and
$U = 0.3 D$.

The top panel shows the standard SET, with no direct tunneling
channel. At the lowest temperature ($k_BT=10^{-9}D$), the model
Hamiltonian close to the strong-coupling fixed point, the Kondo
screening makes the quantum dot transparent to electrons, so that in
the Kondo regime [$\Gamma\ll \min(|\ed|, 2\ed+U)$] the Wiedemann-Franz
law pushes the ratio $3\kappa/\pi^2T$ to the unitary limit
$2e^2/h$. At higher temperatures, the Kondo cloud evaporates and the
thermal conductance drops steepily. The maxima near $\ed=0$ and
$\ed=-U$ reflect the two resonances associated with the transitions
$c_d^1\leftrightarrow c_d^0$ and $c_d^1\leftrightarrow c_d^2$.

In the next two panels, the direct tunneling amplitude substantially
increased, the current through the dot tends to interfere with the
current bypassing the dot. To show that a particle-hole transformation
is equivalent to changing the sign of the amplitude $t$, we compare
the curves with $t=0.16\,D$ (second panel) with $t=-0.16D$ (third
panel). At low temperatures, in the former (latter) case, the
interference between the $c_d^1\leftrightarrow c_d^0$ and
$c_d^1\leftrightarrow c_d^2$ transitions is constructive near $\ed=0$
($\ed=-U$) and destructive near $\ed=U$ ($\ed=0$). At intermediate dot
energies, $\ed\approx -U/2$, the amplitudes for direct transition and
for transition through the dot have orthogonal phases and fail to
interfere, so that the resulting current is the sum of the two
individual currents.

In the bottom panel, the direct tunneling amplitude $t$ is
dominant. For gate potentials disfavoring the formation of a dot
moment, heat flows from one electrode to the other. In the Kondo
regime, however, at low temperatures, the Kondo cloud coupling the dot
to the electrode orbitals closest to it blocks transport between the
electrodes. As the Kondo cloud evaporates, the thermal conductance in
the $\ed\approx-U/2$ rises with temperature, so that the resulting
profile is symmetric to the one in the top panel. The two resonances
near $\ed=0$ and $\ed=-U$, which are independent of Kondo screening,
keep the thermal conductance low even at relatively high temperatures.

\begin{figure}[h]
\begin{center}
\includegraphics[clip=true,width=8.5cm]{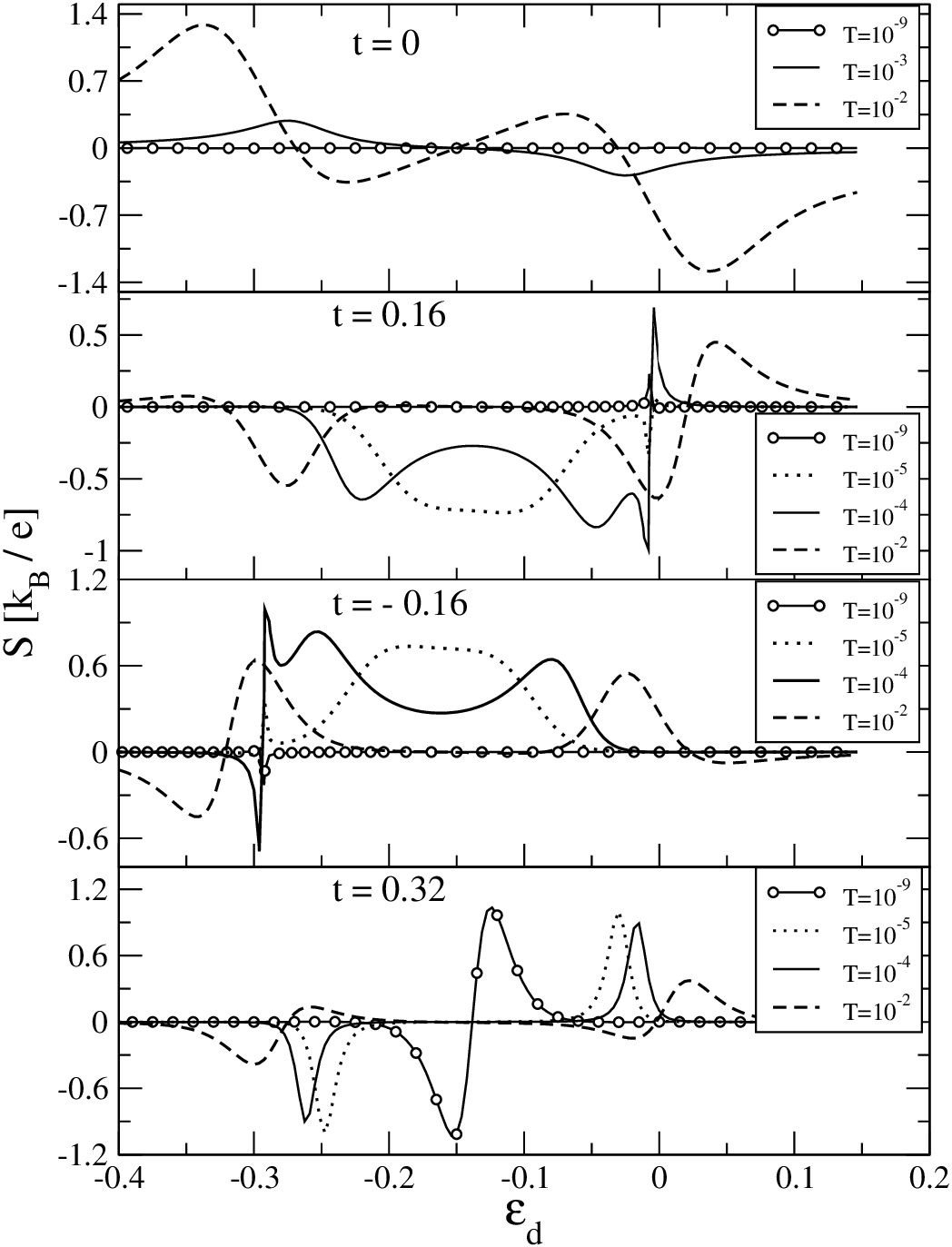}
\end{center}
\caption{Thermopower as a function of $\varepsilon_{d}$ for the four
  tunneling amplitudes $t$ in Fig.~\ref{fig:1}. Again $\Gamma=0.01D$ and $U = 0.3
  D$. For each $t$, the profile of the thermal power is presented at
  the indicated temperatures. \label{fig:2}}
\end{figure}

Figure~\ref{fig:2} shows thermopower profiles for the same amplitudes
$t$ discussed in Fig.~\ref{fig:1}. In contrast with the thermal
conductance, the thermopower is sensitive to particle-hole asymmetry:
heat currents due to holes (electrons) make it positive
(negative). For the standard SET ($t=0$, top panel), the thermopower
is negligible at low temperatures and vanishes at the particle-hole
symmetric parametrical point $\varepsilon_{d} = -U/2$.

With $|t|=0.16D$, particle-hole symmetry is broken at $\ed=-U/2$, and
temperatures comparable to $T_K$ make the thermopower sizeable in the
Kondo regime. For $t=0.16D$, the sensitivity to particle-hole
asymmetry makes the interference between electron (hole) currents
constructive (destructive) for both $\ed=0$ and for $\ed=U$, while for
$t=-0.16D$ it is destructive (constructive).

For $t=0.32D$, direct tunneling again dominant, in the Kondo regime
($\ed\approx -U/2$) the thermopower becomes sensitive to the Kondo
effect, which is chiefly due to electrons (holes) above (below) the
  Fermi level. The thermopower therefore emerges as a
probe of direct-tunneling leaks in SETs, one that may help identify
the source of interference in this and other nanodevices.

\section{Side-coupled quantum dot}
\begin{figure}[h]
\begin{center}
\includegraphics[clip=true,width=8.5cm]{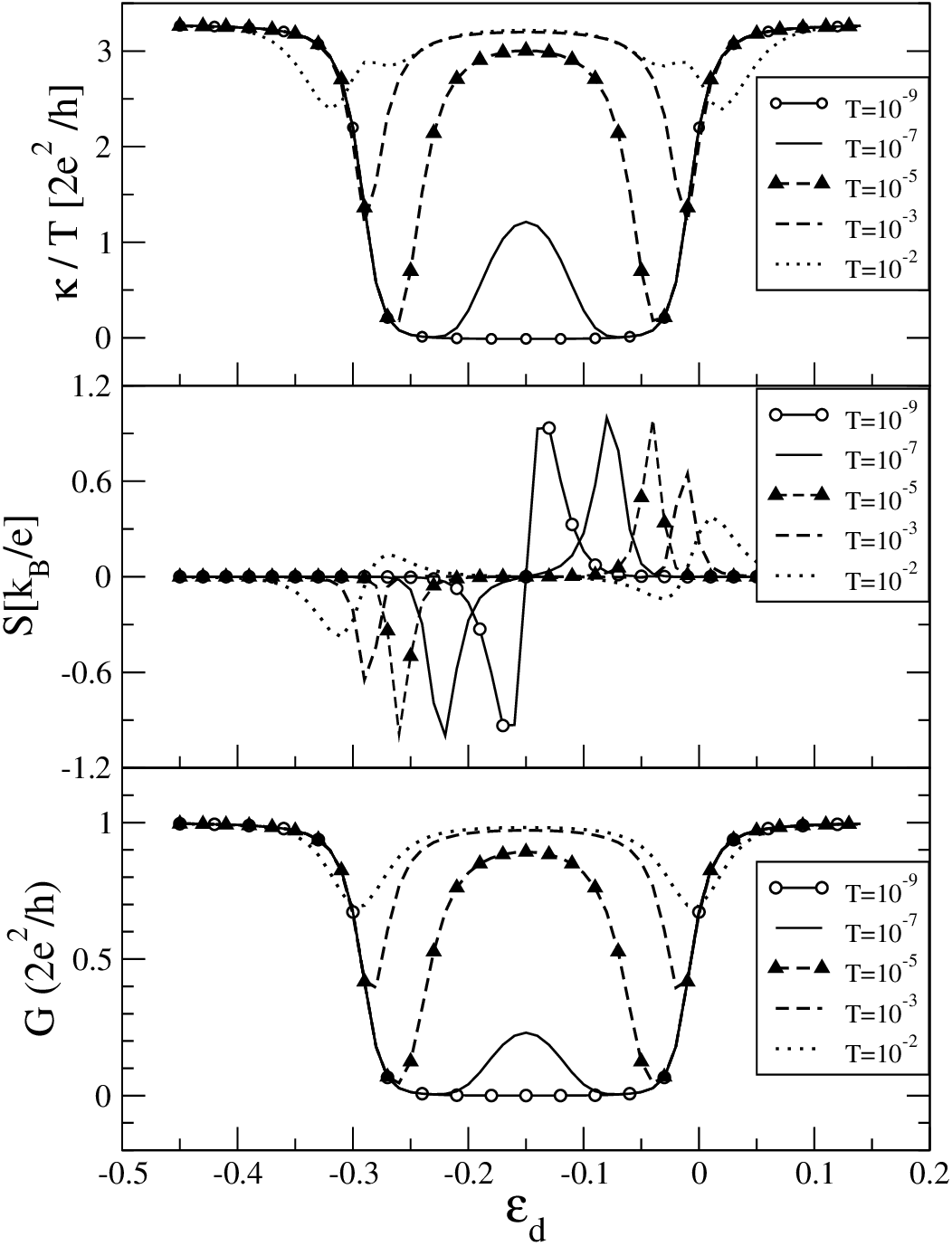}
\end{center}
\caption{The thermal conductance, thermopower and electric conductance
  are presented as a function of $\varepsilon_{d}$.  The temperature
  dependence of each quantity is also presented.\label{fig:3}}
\end{figure}

We have also studied the $T$-shaped device, in which the dot is
side-coupled to \revised{a quantum}{the} wire
\cite{sato}. Again, we considered the Seebeck setup. The
  quantum wire now shunting the two electrodes, we drop the coupling
  proportional to $t$ on the right-hand side of Eq.~(\ref{eq:ham}) and
 employ the standard Anderson Hamiltonian
\begin{eqnarray}
H_s = \sum_{k} \ek c_{k}^{\dag}c_{k}^{\ndag}+
V\sum_{k}( c_{k,\sigma}^{\dag} c_{d}^{\ndag} + H.c.) +
H_{d}
\end{eqnarray}
where $H_{d} = \ed c_{d}^{\dag} c_{d}^{\ndag} + U n_{d,\uparrow}
n_{d,\downarrow}$ is the dot Hamiltonian. The transmission probability
is now given by ${\mathcal 
  T}(\varepsilon,T) = 1+ \pi \rho V^{2} \Im\{G_{d,d}\}(\varepsilon,T)$
where $G_{d,d}(\varepsilon,T)$. Following the procedure outlined
above, we have diagonalized the Hamiltonian $H_s$ iteratively and
computed the electrical conductance, the thermal conductance and the
thermopower as functions of the gate potential
$\varepsilon_{d}$. Figure~\ref{fig:3} displays results for $U = 0.3
D$, and $\Gamma = 0.01\,D$.

Not surprisingly---the wire is equivalent to a large tunneling
amplitude, i.~e.,\ to $t\sim D$---the transport coefficients mimic those of the $t=0.32\,D$ SET.
At low temperatures ($T \ll T_{K}$) in the Kondo regime, for instance,
the Kondo cloud blocks transport through the wire segment closest to
the dot.  The thermal and electrical conductances thus vanish for
$\ed\approx-U/2$. As the temperature rises, the evaporation of the
Kondo cloud allows transport and both conductances rise near the
particle-hole symmetric point. At low temperatures, the sensitivity to
particle-hole asymmetry enhances the thermopower in the Kondo regime,
a behavior analogous to the bottom panel in Fig.~\ref{fig:2}.

\section{Conclusions}
We have calculated the transport coefficients for the SET and the
side-coupled geometries. In both cases, the thermal dependence and the
gate-voltage profiles show signatures of the Kondo effect and of
quantum interference. Our essentially exact NRG results identify
trends that can aid the interpretation of experimental results. In the
side-coupled geometry, in particular, the Kondo cloud has marked
effects upon the thermopower.

\section{Acknowledgments}
Financial support by the FAPESP and the CNPq is acknowledged.

\end{document}